# Combinational Logic Circuit Design with the Buchberger Algorithm

Germain Drolet

Department of Electrical & Computer Engineering,

Royal Military College of Canada,

P.O. Box 17000, Station Forces,

Kingston, Ontario, CANADA

K7K 7B4

Tel: (613) 541-6000, extension: 6192

Fax:(613) 544-8107

Email: drolet-g@rmc.ca

**Abstract:** We detail a procedure for the computation of the polynomial form of an electronic combinational circuit from the design equations in a truth table. The method uses the Buchberger algorithm rather than current traditional methods based on search algorithms. We restrict the analysis to a single output, but the procedure can be generalized to multiple outputs. The procedure is illustrated with the design of a simple arithmetic and logic unit with two 3-bit operands and two control bits.

## Section 1: Introduction

A memoryless combinational logic function is a mapping $\mathbf{F}_2^l \to \mathbf{F}_2^m$ from input variables $y_1$, …, $y_l$ to output variables $z_1$, …, $z_m$ where $\mathbf{F}_2 \cong \mathrm{GF}(2)$ denotes the binary field. By *memoryless* we mean that the values of the output variables depend only on the present values of the inputs; previous (or future) values of the input values are irrelevant. A combinational logic function is implemented in an electronic circuit by interconnecting asynchronous logic gates such as NAND, OR, XOR, ... [16]. Each output $z_i$ of a combinational logic function can be expressed as the evaluation of a binary multivariate polynomial of the $l$ input variables $y_1$, …, $y_l$, that we refer to as *polynomial form* of a combinational logic function or truth table. Other authors refer to them as *modulo 2 sum-of-products* (mod-2 SOP) [1, 5, 13], *Reed-Muller representations* [14, 15], *Reed-Muller polynomials* [2], *Reed-Muller algebraic form* [6] or *Reed-Muller (Exclusive-OR) expansions with positive polarities* [8]. A truth table always admits at least one polynomial form. Indeed a sum of product Boolean function of the truth table can always be found [16] and a polynomial form can be derived from the sum of product Boolean function by replacing the logic complements and the OR operations as follows:

(1)
$$\overline{x} = x + 1$$
$$x \vee y = x + y + xy$$

for every $x, y \in \mathbf{F}_2$. Polynomial forms are often desirable because they lead to well structured easily testable designs [12]. In addition the implementations use XOR and AND gates only, of which very efficient implementations are obtained in the *Pass Transistor Logic* technology (used in low-power dissipation applications) [1, 5, 13].

In the early phases of design, a combinational logic function is described by a *truth table*. Each row of the truth table is a set of equations (Boolean, sum-of-product (SOP), polynomial) in terms of the input variables, output variables (and possibly some intermediate variables to be eliminated), the solutions of which specify the values of the output variables for some values of the input variables. In general the input variable values specified by the truth table are restricted to a set $S \subset \mathbf{F}_2^l$; outputs corresponding to inputs lying in $\mathbf{F}_2^l - S$ are not specified and are called *don't cares*. As a result the truth table does not specify a unique combinational logic function and a designer will select the combinational logic function having the smallest complexity. We present a method that allows one to compute a simple polynomial form of a combinational logic function from the equations of a truth table when $m = 1$, i.e. combinational logic functions $\mathbf{F}_2^l \to \mathbf{F}_2$. The restriction $m = 1$ simplifies the notations and some arguments; the results can easily be generalized to $m > 1$. When all equations of the truth table used to describe any output variable do not involve any of the other output variables then the mapping $\mathbf{F}_2^l \to \mathbf{F}_2^m$ can be synthesized as $m$ mappings $\mathbf{F}_2^l \to \mathbf{F}_2$ (the problem of reusability of common terms is much more difficult and is not addressed in the paper). We assume that the reader is familiar with elementary concepts of electronic synthesis of combinational logic functions [16] and with an elementary knowledge of computational algebraic geometry [4].

Traditionally a truth table is simplified and an electronic circuit synthesized using Karnaugh maps [16]. Karnaugh maps ultimately yield the circuit with the smallest complexity but the method is rarely used since it quickly becomes impractically complex as the number of variables increases. A recently proposed practical method with a larger number of variables utilizes the Walsh Transform [11]. The procedure leads to the Reed-Muller form with mixed polarities (the variables may appear in the positive and/or negative polarities) and a minimal number of product terms. The procedure does not allow *don't care* values since it requires that the truth table be completely specified. Combinational logic functions are more often obtained by search algorithms using *Boolean decision diagrams* [9]. The search algorithm does not necessarily find the least complex circuit and can be very sensitive to the ordering of the input variables, especially for the resolution of *don't care* values. The proposed method uses the Buchberger algorithm and thus calculates rather than search for the polynomial form of a combinational logic function. The algorithm guarantees that the polynomial form of the combinational logic function obtained has the smallest multidegree with respect to the monomial ordering used, of all polynomial form representations of all combinational logic functions for any resolution of the *don't cares*. In section 2 we present the basic approach of the solution, in section 3 we consider the practical computation aspects which are further illustrated on a simple example in section 4. Section 5 presents some possible areas of future development.

## Section 2: General Solution

In the following $y_1$, …, $y_l$ denote the $l$ input variables of a combinational logic function with output variable $z$ and $x_1$, $x_2$, …, $x_k$ are $k \geq 0$ intermediate variables. The notation $x_{k+1} = z$, $x_{k+2} = y_1$, ..., $x_n = y_l$ ($n = k + l + 1$) will be used as it simplifies the expressions. We denote by $\mathbf{F}_2[x_1, x_2, ..., x_n]$ the multivariate polynomial ring with coefficients in the binary field $\mathbf{F}_2 \cong \mathrm{GF}(2)$. Without loss of generality we assume that the equations in each row of the truth table are written in the form of homogeneous polynomial equations (we assume that the reader is familiar with *truth tables* and rather than giving a formal definition we present an example below). Let $s$ denote the number of rows in the truth table, $t_i$ the number of equations in row $1 \leq i \leq s$ of the table and $f_{ij} = 0$, $i = 1, ..., s$, $j = 1, ..., t_i$ the $j$-th homogeneous polynomial equation in row $i$ of the truth table. $f_{ij}$ is a binary multivariate polynomial, $f_{ij} \in \mathbf{F}_2[x_1, x_2,..., x_n]$, and the variety $V_i = V(f_{i,1}, f_{i,2},..., f_{i,t_i}) \subset \mathbf{F}_2^n$ is the set of solutions of the equations in row $i$. The truth table specifies a variety $V_F = \bigcup_{i=1}^{s} V_i$, where $V_i$ specifies the value of the output variable $z = x_{k+1}$ for some values of the input variables $x_{k+2}, ..., x_n$.

*Example:* Table 1 shows the truth table of an arithmetic and logic unit (ALU) operating on two 3-bit words representing the binary expansion of unsigned integers between 0 and 7. The truth table uses 8 binary input variables ($OP_0, OP_1, a_0, a_1, a_2, b_0, b_1, b_2$), 4 output variables ($c_0, c_1, c_2$, *flag*) and 2 intermediate variables ($carry_0, carry_1$). The values of the input bits $OP_0$ and $OP_1$ determine one of three possible operations performed by the ALU on the 3 bit-words. The meaning of the output variable *flag* depends on the operation performed:

| $OP_0$ | $OP_1$ | Operation | *flag* |
|---|---|---|---|
| 0 | 1 | largest | equal |
| 1 | 0 | smallest | equal |
| 1 | 1 | sum | overflow |

In section 4 we compute the polynomial form of the combinational logic function of the output variable $c_2$.

Returning to the general solution, we consider the projection maps:

(2)
$$\begin{aligned} \pi_l: \quad & \mathbf{F}_2^n \quad \to \quad \mathbf{F}_2^{n-l} \\ \pi_l: \quad & (a_1,...,a_n) \mapsto (a_{l+1}, a_{l+2},...,a_n) \end{aligned}$$

$l = 1, 2, ..., n-1$. Since the polynomials $f_{ij}$, $i = 1, 2, ..., s$, $j = 1, 2, ..., t_i$ represent a well-defined mapping with input variables $x_{k+2}, ..., x_n$ and output variable $x_{k+1}$, the projection map:

(3)
$$\begin{aligned} v: \quad & \pi_k(V_F) \to S = \pi_{k+1}(V_F) \\ v: \quad & (a_{k+1},...,a_n) \mapsto (a_{k+2},...,a_n) \end{aligned}$$

must be bijective. Consequently the function $x_{k+1} = \tau(x_{k+2}, x_{k+3},..., x_n)$ determined by the truth table is given by the first component of $v^{-1}(s)$, $\forall s \in S \subset \mathbf{F}_2^l$. Also $\pi_{k+1}(V_F) = S$ is the domain of the truth table. In the following we let $>$ denote a monomial ordering on $x_1, x_2,..., x_n$ which is of $k$ and $k+1$ elimination type.

**Definition:** A *binary multivariate polynomial form of the truth table* (or simply *polynomial form*) is a polynomial $f \in \mathbf{F}_2[x_{k+1}, x_{k+2},..., x_n]$ such that:
  (i) $V(f) \supset \pi_k(V_F)$,
  (ii) $\mathrm{LT}(f) = x_{k+1}$ (with respect to $>$).

The restriction of any polynomial form of the truth table to $S$ is clearly equal to the function $\tau$. Define the radical ideal $I = I(V_F) \subset \mathbf{F}_2[x_1, x_2,..., x_n]$ and the $k$-th elimination ideal:

(4) $$I_k = I \cap \mathbf{F}_2[x_{k+1}, x_{k+2},..., x_n].$$

$I_k$ is a radical ideal in the polynomial ring $\mathbf{F}_2[x_{k+1}, x_{k+2},..., x_n]$ containing all polynomial forms of the truth table. Let $G = \{g_1, g_2,..., g_t\} \subset \mathbf{F}_2[x_1, x_2,..., x_n]$ be a reduced Gröbner basis of $I = I(V_F)$ and without loss of generality let $G_k = G \cap \mathbf{F}_2[x_{k+1}, x_{k+2},..., x_n] = \{g_1, g_2,..., g_r\}$ where $r \leq t$. $G_k$ is a reduced Gröbner basis of $I_k$, since $>$ is of $k$-elimination type. In addition $>$ is of $k+1$-elimination type and since $I_k$ contains all polynomial forms of the truth table then $x_{k+1} \in \langle \mathrm{LT}(I_k) \rangle$. It follows that $\mathrm{LT}(g_j) = x_{k+1}$ for some $1 \leq j \leq r$. But $G_k$ is reduced (and therefore minimal) so that there can only be one such polynomial in $G_k$, $G_{k+1} = \{g_1,..., g_{j-1}, g_{j+1},..., g_r\}$. Finally, noticing that $g_j$ vanishes on $V_F$ we conclude that $g_j$ is a polynomial form of the truth table. The results are summarized in the following:

**Theorem 1:** *Let $I_k$ be the $k$-th elimination ideal of $I(V_F)$ and $G_k$ a reduced Gröbner basis of*

$I_k$. Then $I_k$ contains all the polynomial forms of the truth table and $G_k$ contains exactly one polynomial form of the truth table.

We expect 1 to be the power of every variable $x_i$, $i = k+2$, ..., $n$, in the monomials of $g_j$. Indeed, $x_i^2 + x_i \in I$, $\forall i = 1, 2, ..., n$ since $V_F \subset \mathbf{F}_2^n$ and it follows that

(5) $$x_i^2 \in \langle \text{LT}(I_{k+1}) \rangle = \langle \text{LT}(g_1), ..., \text{LT}(g_{j-1}), \text{LT}(g_{j+1}), ..., \text{LT}(g_r) \rangle,$$

$\forall i = k+2, k+3, ..., n$, which means that:

(6) $$\langle x_{k+2}^2, ..., x_n^2 \rangle = \langle \text{LT}(g_1), ..., \text{LT}(g_{j-1}), \text{LT}(g_{j+1}), ..., \text{LT}(g_r) \rangle.$$

The result then follows from $G_k$ being reduced which means that none of the above leading terms divide any of the monomials of $g_j$. We also expect $g_j$ to have a *relatively simple expression* [4].

## Section 3: Practical Computation

In most cases, the ideal $I = I(V_F)$ can be computed with the method described below. The conditions that the truth table must satisfy in order for the method to apply are detailed. Let $\overline{\mathbf{F}}_2 \supset \mathbf{F}_2$ denote an algebraically closed extension of $\mathbf{F}_2$. In addition to the previously defined $V_i = V(f_{i,1}, f_{i,2}, ..., f_{i,t_i}) \subset \mathbf{F}_2^n$ for the $i$-th row of the truth table, we introduce:

(7) $$\overline{V}_i = V(f_{i,1}, f_{i,2}, ..., f_{i,t_i}) \subset \overline{\mathbf{F}}_2^n,$$

(8) $$I_i = \langle f_{i,1}, f_{i,2}, ..., f_{i,t_i}, x_1^2 + x_1, ..., x_n^2 + x_n \rangle \subset \mathbf{F}_2[x_1, x_2, ..., x_n],$$

(9) $$W_i = V(I_i) \subset \mathbf{F}_2^n,$$

(10) $$\overline{W}_i = V(I_i) \subset \overline{\mathbf{F}}_2^n,$$

(11) $$I' = \prod_{i=1}^{s} I_i \subset \mathbf{F}_2[x_1, x_2, ..., x_n].$$

Clearly $\overline{W}_i = W_i = V_i \subset \overline{V}_i$. A basis of $I'$ can easily be computed, it being a product of ideals. We will see that the ideals $I_i$, $i = 1, 2, ..., s$, are radical ideals and that a typical truth table often yields ideals $I_i$ that are pairwise coprime. The practical interest in the ideal $I'$ is emphasized by the following theorem.

**Theorem 2:** *If the ideals $I_i$ are radical and pairwise coprime then $I' = I(V_F) = I$.*

*proof:* When the $I_i$'s are pairwise coprime $\bigcap_{i=1}^{s} I_i = \prod_{i=1}^{s} I_i = I'$. It follows that:

$$V(I') = V\left(\bigcap_{i=1}^{s} I_i\right)$$
$$= \bigcup_{i=1}^{s} V(I_i)$$
$$= \bigcup_{i=1}^{s} W_i$$
$$= \bigcup_{i=1}^{s} V_i$$
$$= V_F$$

and all zeroes of $I'$ in $\overline{\mathbf{F}}_2^n$ are confined to $V_F \subset \mathbf{F}_2^n$. It follows from the Nullstellensatz [10] and the definitions that:

(12) $$I = I(V_F) = I(V(I')) = \sqrt{I'}.$$

If moreover all $I_i$'s are radical ideals in $\mathbf{F}_2[x_1, x_2, ..., x_n]$ then $I' = \bigcap_{i=1}^{s} I_i$ is also a radical ideal [4]:

$\sqrt{I'} = I'.$ ∎

**Proposition 1:** *The ideals $I_i = \langle f_{i,1}, f_{i,2}, ..., f_{i,t_i}, x_1^2 + x_1, ..., x_n^2 + x_n \rangle$, $i = 1, 2, ..., s$ are radical.*

*proof*: $\overline{W}_i = W_i \subset \mathbf{F}_2^n$ is obviously a finite variety. It follows that the ideal $I_i$ is zero-dimensional [3]. The result then follows from Seidenberg's lemma [3]. ∎

**Proposition 2:** *$I_i$ and $I_j$ are coprime if and only if $W_i \cap W_j = \varnothing$. In particular, if all the rows of the truth table correspond to pairwise disjoint sets of input variable values then the ideals $I_i$ are pairwise coprime.*

*proof*: It is well known that $W_i \cap W_j = V(I_i) \cap V(I_j) = V(I_i + I_j)$. If $I_i$ and $I_j$ are coprime then $I_i + I_j = \langle 1 \rangle$ and $V(\langle 1 \rangle) = \varnothing$. For the converse, we recall that all zeroes of $I_i$ and $I_j$ in $\overline{\mathbf{F}}_2^n$ are confined to $\mathbf{F}_2^n$ and so if $V(I_i + I_j) = \varnothing$ then by the *weak* Nullstellensatz [10] we have $I_i + I_j = \langle 1 \rangle$, i.e. the ideals $I_i$ and $I_j$ are coprime.

For the second part of the statement we first notice that $\pi_{k+1}(W_i)$ is the set of input variable values of row $i$ of the truth table. Next, we easily see that $\pi_{k+1}(W_i \cap W_j) \subset \pi_{k+1}(W_i) \cap \pi_{k+1}(W_j)$. So if $\pi_{k+1}(W_i) \cap \pi_{k+1}(W_j) = \varnothing$ then $\pi_{k+1}(W_i \cap W_j) = \varnothing$ and the only way in which this can happen is if $W_i \cap W_j = \varnothing$. The result then follows from the first part of the statement. ∎

In a typical situation the rows of the truth table (possibly after some rearrangement) correspond to pairwise disjoint sets of values of the input variables $x_{k+2}$, …, $x_n$. Generators of $I'$ defined in equations (8), (11) are easily found and by theorem 2, $I = I(V_F) = I'$. The Buchberger algorithm with respect to the lexicographic monomial ordering gives Gröbner basis $G$ from which $G_k$ is easily obtained. The procedure is illustrated in the example that follows.

## Section 4: Example

Referring to table 1, we illustrate the application of the procedure by computing the polynomial form of the truth table for the output variable $c_2$ (the polynomial form for any of the output variables can be obtained similarly). In this example we simply disregard the three other output variables (in general some or all of those output variables may be used as intermediate variables). We start by forming the ideals $I_1$, $I_2$, ..., $I_{16} \subset \mathbf{F}_2[carry_0, carry_1, c_2, OP_0, OP_1, a_2, a_1, a_0, b_2, b_1, b_0]$ for each of the 16 rows of the truth table. The following ideals are given as an illustration:

(13)
$$I_{15} = \langle OP_0 + 1, OP_1, a_2 + b_2, a_1 + b_1, a_0 + b_0, c_2 + a_2, carry_0^2 + carry_0$$
$$, carry_1^2 + carry_1, c_2^2 + c_2, OP_0^2 + OP_0, OP_1^2 + OP_1, a_2^2 + a_2, a_1^2 + a_1, a_0^2 + a_0$$
$$, b_2^2 + b_2, b_1^2 + b_1, b_0^2 + b_0 \rangle$$

(14)
$$I_2 = \langle OP_0, OP_1 + 1, a_2 + 1, b_2, c_2 + a_2, carry_0^2 + carry_0$$
$$, carry_1^2 + carry_1, c_2^2 + c_2, OP_0^2 + OP_0, OP_1^2 + OP_1, a_2^2 + a_2$$
$$, a_1^2 + a_1, a_0^2 + a_0, b_2^2 + b_2, b_1^2 + b_1, b_0^2 + b_0 \rangle$$

(15)
$$I_1 = \langle OP_0 + 1, OP_1 + 1, c_2 + a_2 + b_2 + carry_1$$
$$, carry_1 + a_1 b_1 carry_0 + (a_1 b_1 carry_0 + 1)(a_1 b_1 + a_1 carry_0 + b_1 carry_0)$$
$$, carry_0 + a_0 b_0, carry_0^2 + carry_0, carry_1^2 + carry_1, c_2^2 + c_2, OP_0^2 + OP_0$$
$$, OP_1^2 + OP_1, a_2^2 + a_2, a_1^2 + a_1, a_0^2 + a_0, b_2^2 + b_2, b_1^2 + b_1, b_0^2 + b_0 \rangle$$

As stated in proposition 1 the above ideals are all radical. In addition we see that all the rows of the truth table correspond to pairwise disjoint sets of input variables and by proposition 2 the ideals are pairwise coprime. It follows from theorem 2 that $I = I' = \prod_{i=1}^{16} I_i$ and its generators are obtained by taking the product of the set of generators of $I_1$, $I_2$, ..., $I_{16}$. The Buchberger algorithm (*SINGULAR* [7]) is used to calculate the reduced Gröbner basis $G$ of $I' = I$ which contains 4 polynomials. $G_2 = G \cap \mathbf{F}_2[c_2, OP_0, OP_1, a_2, a_1, a_0, b_2, b_1, b_0]$ generates the second elimination ideal $I_2$ of $I$ and contains 2 polynomials:

$$G_2 = \{OP_0 OP_1 + OP_0 + OP_1 + 1, c_2 + OP_1 a_0 b_0 b_1 + OP_1 a_1 b_1 + OP_1 a_2$$
(16)
$$+ OP_1 a_0 b_0 a_1 + OP_1 a_2 b_2 + OP_1 b_2 + OP_0 a_1 b_1 + OP_0 a_0 b_0 b_1$$
$$+ OP_0 a_0 b_0 a_1 + OP_0 a_2 b_2 + a_0 b_0 b_1 + a_1 b_1 + a_0 b_0 a_1\}$$

By theorem 1 the second elimination ideal of $I$ contains all polynomial forms of the truth table one of which is contained in $G_2$. As mentioned the power of every variable is 1 and a *relatively simple* polynomial form expression for $c_2$ in terms of the input variables is then:

(17)
$$c_2 = OP_1 a_0 b_0 b_1 + OP_1 a_1 b_1 + OP_1 a_2 + OP_1 a_0 b_0 a_1 + OP_1 a_2 b_2$$
$$+ OP_1 b_2 + OP_0 a_1 b_1 + OP_0 a_0 b_0 b_1 + OP_0 a_0 b_0 a_1 + OP_0 a_2 b_2 + a_0 b_0 b_1$$
$$+ a_1 b_1 + a_0 b_0 a_1$$

Out of all the polynomial form expression for $c_2$ that meet all specified values of the truth table (and filling the unspecified *don't care* values in some way), the above expression has the smallest multidegree with respect to the lexicographic order chosen.

The author has recently verified that all equations of the truth table can simultaneously be solved, thus obtaining the polynomial forms of the $m > 1$ combinational logic functions in one step. The theoretical justification of the applicability of the method is a simple generalization of the results presented in this paper. Although the run-time has not been thoroughly analyzed, nor compared with other methods, we report that the simultaneous computation of all 4 outputs of the 3-bit ALU takes 13 seconds on a personal computer with a 1.5 GHz *AMD Athlon* processor with 512 MB of RAM.

*Note:* The result is more efficiently computed by taking the products of the set of generators of $I_1$, $I_2$, …, $I_{16}$ two by two and calculating a Gröbner basis of each intermediate product.

## Section 5: Possible areas of future development

Search algorithms based on a Boolean decision diagrams as well as the procedure that we develop have the disadvantage that the complexity of the resulting polynomial form can vary considerably with the ordering in which the input variables are treated. We conjectured that this difficulty can be resolved for our procedure by using a different monomial ordering of the variables $x_1$, $x_2$, …, $x_n$ as follows: a *product order* consisting of the lexicographic order for $x_1$, $x_2$, …, $x_{k+1}$ together with the graded lexicographic order for the input variables $x_{k+2}$, $x_{k+3}$, …, $x_n$ would satisfy the requirements of the procedure and remove the dependency of the order in which the input variables are listed. This is currently under investigation.

This paper demonstrates the applicability of the Buchberger algorithm to the calculation of a simple *polynomial form* of a combinational logic function possibly containing *don't care* values. A simple example (3-bit ALU) has been presented. A thorough analysis of the run time and comparison with other methods over a larger set of common benchmarks is the subject of ongoing research, the results of which will be made available soon.

The reusability of common terms is a much more difficult problem which has not been considered in the paper. It is the author's opinion that the method presented will not solve this complex problem.

**References:**

1. Al-Khalili D., Kechichian K., Liqin Dong, "Synthesis of low power CMOS digital circuits based on fully restored pass logic", *Canadian Conference on Electrical and Computer Engineering, 1996*, Vol. 1, 142 - 145 (1996)

2. Almaini A.E.A., Thomson P., Hanson D., "Tabular techniques for Reed-Muller logic", *Int. J. Electronics*, vol. 70, no. 1, 23 - 34, (1991)

3. Becker T., Weispfenning V., *Gröbner Bases*, GTM, Springer-Verlag 1993

4. Cox D., Little J., O'Shea D., *Ideals, Varieties, and Algorithms*, 2nd Edition, UTM, Springer 1997

5. Gallant M., Al-Khalili D., "Synthesis of low power circuits using combined pass logic and CMOS topologies", *Proceedings of the Tenth International Conference on Microelectronics, ICM 1998*, 59 - 62 (1998)

6. Green D.H., Khuwaja G.A., "Simplification of switching functions expressed in Reed-Muller algebraic form", *IEE Proceedings part E*, vol. 139, no. 6, 511 - 518, (1992)

7. Greuel G.-M., Pfister G., Schoenemann H., *SINGULAR: A Computer Algebra System for Polynomial Computations*, FB Mathematik der Universitaet, D-67653 Kaiserslautern, version 2-0-4, April 2003 (http://www.singular.uni-kl.de/).

8. Habib M.K., "New Approach for the generation of minimal Reed-Muller Exclusive-OR expansions with mixed polarity", *Int. J. Electronics*, vol. 66, no. 6, 865 - 874, (1989)

9. Hansen J.P., Sekine M., "Synthesis by spectral translation using Boolean Decision Diagrams", *33rd Design Automation Conference 1996*, 248 - 253 (1996)

10. Matsumura H., *Commutative Ring Theory*, Cambridge University Press 1989

11. Porwik, P., "Efficient Calculation of the Reed-Muller Form by Means of the Walsh Transform", *Int. J. Appl. Math. Comput. Sci.*, vol. 12, no. 4, 571- 579, (2002)

12. Reddy S.M., "Easily testable realizations for logic functions", *IEEE Trans. on Computers*, vol. C-21, no. 11, 1183 - 1188, (1972)

13. Saleem D., Al-Khalili D., "Low power conditional sum adder using pass logic topology", *IEEE Canadian Conference on Electrical and Computer Engineering 1998*, Vol. 1, 9 - 12 (1998)

14. Saul J.M., "An algorithm for the multi-level minimization of Reed-Muller representations", *Proceedings of the IEEE International Conference on Computer Design, October 1991*, 634 - 637 (1991)

15. Saul J.M., "Logic synthesis for arithmetic circuits using the Reed-Muller representation", *Proceedings of the IEEE International Conference on Computer Design, October 1992*, 109 - 113 (1992)

16. Winkel D., Prosser F., *The Art of Digital Design*, Prentice-Hall 1980

| Input variables | | | | | | | | Output variables | | | |
|---|---|---|---|---|---|---|---|---|---|---|---|
| $OP_0$ | $OP_1$ | $a_2$ | $a_1$ | $a_0$ | $b_2$ | $b_1$ | $b_0$ | $c_2$ | $c_1$ | $c_0$ | flag |
| 1 | 1 | X | X | X | X | X | X | $a_2 + b_2 + carry_1$ | $a_1 + b_1 + carry_0$ $carry_1 = ...$ | $a_0 + b_0$ $carry_0 = a_0 b_0$ | $flag = \overline{a_2} \overline{b_2} carry_1$ $\vee \overline{a_2} b_2 \overline{carry_1}$ $\vee a_2 \overline{b_2} \overline{carry_1}$ $\vee a_2 b_2 \overline{carry_1}$ |
| 0 | 1 | 1 | X | X | 0 | X | X | $a_2$ | $a_1$ | $a_0$ | 0 |
| 0 | 1 | 0 | X | X | 1 | X | X | $b_2$ | $b_1$ | $b_0$ | 0 |
| 0 | 1 | $b_2$ | 1 | X | $a_2$ | 0 | X | $a_2$ | $a_1$ | $a_0$ | 0 |
| 0 | 1 | $b_2$ | 0 | X | $a_2$ | 1 | X | $b_2$ | $b_1$ | $b_0$ | 0 |
| 0 | 1 | $b_2$ | $b_1$ | 1 | $a_2$ | $a_1$ | 0 | $a_2$ | $a_1$ | $a_0$ | 0 |
| 0 | 1 | $b_2$ | $b_1$ | 0 | $a_2$ | $a_1$ | 1 | $b_2$ | $b_1$ | $b_0$ | 0 |
| 0 | 1 | $b_2$ | $b_1$ | $b_0$ | $a_2$ | $a_1$ | $a_0$ | $a_2$ | $a_1$ | $a_0$ | 1 |
| 1 | 0 | 1 | X | X | 0 | X | X | $b_2$ | $b_1$ | $b_0$ | 0 |
| 1 | 0 | 0 | X | X | 1 | X | X | $a_2$ | $a_1$ | $a_0$ | 0 |
| 1 | 0 | $b_2$ | 1 | X | $a_2$ | 0 | X | $b_2$ | $b_1$ | $b_0$ | 0 |
| 1 | 0 | $b_2$ | 0 | X | $a_2$ | 1 | X | $a_2$ | $a_1$ | $a_0$ | 0 |
| 1 | 0 | $b_2$ | $b_1$ | 1 | $a_2$ | $a_1$ | 0 | $b_2$ | $b_1$ | $b_0$ | 0 |
| 1 | 0 | $b_2$ | $b_1$ | 0 | $a_2$ | $a_1$ | 1 | $a_2$ | $a_1$ | $a_0$ | 0 |
| 1 | 0 | $b_2$ | $b_1$ | $b_0$ | $a_2$ | $a_1$ | $a_0$ | $a_2$ | $a_1$ | $a_0$ | 1 |
| 0 | 0 | X | X | X | X | X | X | X | X | X | X |
| X = *don't care* | | | | | | | | | | | |

**Table 1:** truth table of a simple 3-bit arithmetic and logic unit